\documentclass[aps,showpacs,superscriptaddress]{revtex4}
\usepackage{bm}
\usepackage{epsfig}
\usepackage{times}
\usepackage{amssymb}
\usepackage{amsmath}
\usepackage{natbib}
\usepackage{color}
\usepackage[latin1]{inputenc}
\usepackage[loose,nice]{units}       

\begin{document}
\title{Formulae for partial widths derived from the Lindblad equation}
\author{S{\o}lve Selst{\o}}
\affiliation{Oslo and Akershus University College for Applied Sciences, N-0130
Oslo, Norway}


\begin{abstract}
A method for calculating partial widths of auto-ionizing states is proposed. It combines either a complex absorbing potential or exterior complex scaling with the Lindblad equation. The corresponding classical rate equations are reproduced, and the trace conservation inherent in the Lindblad equation ensures that the partial widths sum up to the total width of the initial auto-ionizing state.

\end{abstract}

\maketitle

\section{Introduction}
Resonances, or meta-stable states, play a crucial role in several physical phenomena -- ranging from solid state physics \cite{Pan1994,Sajeev2008,Ferron2009,Genkin2010} and plasma physics \cite{Jacobs2001,Jacobs2009}
via atomic and molecular physics \cite{Lindroth1994,Martin1989,Moiseyev1978,QuantChem2011} to nuclear physics \cite{Okolowicza2003,Diaz-Torres2008}.
The population of such states are frequently assumed to follow an exponential decay law. However, we know that this ``law'' is broken -- both at short and long time scales. The former is due to the quantum Zeno paradox, whereas the latter is related to
the fact that there exists a lower threshold for the energy of the system
\cite{Sakurai1994}.
In a large intermediate time region, however, exponential decay remains a perfectly valid approximation.
In such a context it is customary to assume that as the population of some initial meta-stable state decreases exponentially, also the population of the (possibly) various decay products follow rate equations. When making this assumption certain
interference effects are neglected, and one may say that this is a
semi-classical assumption. This suggests that an approach based on a master
equation describing a density matrix rather than a pure state is appropriate \cite{Agarwal1982,Antonenko1994,Jacobs1994}.

Resonance states may appear as solutions of the time-independent Schrödinger equation if we lift the restriction that the solutions should be square integrable and impose outgoing boundary conditions. Asymptotically, these outgoing waves are characterized by complex wave numbers. Consequently, also the eigen energy becomes complex. This is not in violation of the Hermicity of the Hamiltonian since Hermicity only applies to the space consisting of {\it normalizable} wave functions; these complex energy solutions diverge. This is obviously a rather undesirable feature if we want to consider resonance states explicitly. To this end, methods which involve explicitly non-Hermitian terms in the Hamiltonian have proven very useful. When it comes to resonances, the usefulness lies in the fact that such terms are able to ``kill off'' the divergence such that also resonance states may be represented by normalizable functions.
Such explicit non-Hermicity may be introduced in various
ways. It may, e.g., arise by introducing an artificial imaginary potential which is zero in some interior region and increasing towards some boundary (a
complex absorbing potential -- CAP) \cite{Manolopoulos2002,Riss1993,Santra2006}.
Other frequently used techniques are (uniform) complex scaling \cite{Aguilar1971,Balslev1971,Simon1972}, exterior complex scaling (ECS) \cite{Simon1979,Turner1982} and smooth exterior complex scaling \cite{Rom1990,Elander1998}.

In literature, several examples of combining formalism for open quantum systems with explicitly non-Hermitian Hamiltonians are found, see, e.g., \cite{Genkin2008,Caban2005,Bertlmann2006,Selsto2010,Diaz-Torres2010,Selsto2011}.
The term ``open'' in regard to a quantum system indicate that the system is interacting with some environment. As the environment typically has a large number of degrees of freedom, a full solution of the total composite system can usually not be obtained. In such a context, it may be desirable to express the evolution of the smaller system alone where the effect of the environment on the smaller system is included somehow. This is typically achieved by invoking the Born-Markov approximation \cite{Breuer2002}.
The resulting equation of motion, which provides the dynamics of the reduced density matrix of the system, is frequently used in order to describe, e.g., relaxation due to the interaction with a radiation field \cite{Mollow1975,Jacobs2009}, a damped harmonic oscillator \cite{Isar1999} or quantum transport \cite{Harbola2006,Roden2009}.
Moreover, applications in which the system's own continuum is considered to be an environment are also found in literature \cite{Agarwal1984,Genkin2008,Diaz-Torres2010}. (The terms ``open'' and ``unbound'' are sometimes used interchangeably.)
In \cite{Caban2005,Bertlmann2006} it was demonstrated how the Lindblad equation may be used to describe the dynamics of spontaneously decaying particles, i.e. to systems in which the number of particles is not constant.
It has proven useful to combine these ideas with the formalism of second quantization \cite{Harbola2006,Selsto2010}.
Master equations have also proven useful to incorporate the process of spontaneous decay in plasma physics; by introducing the relevant rates into the master equation, non-unitary processes involving auto-ionizing states may treated on equal footing with unitary processes involving radiation and collisions \cite{Jacobs2001,Jacobs2009}.

This work aims to combine non-Hermitian quantum mechanics with a master equation
in order to describe the population dynamics of a system undergoing spontaneous decay to a system with fewer particles. We will make use of the ideas presented in \cite{Selsto2010}, which demonstrates how particle loss due to a CAP may be described in a consistent manner.
This formalism allows us to disregard escaping particles and focus upon whatever is left. (In general this cannot be done using the Schr{\"o}dinger equation.)
Thus, we should be able to maintain information about the part of
the system that remains bound. It is argued that the Lindblad equation is the proper framework for achieving this.
Thus, we take the Lindblad equation to be our starting point; the equation of motion is not derived from any microscopical arguments.
The remainder of the system after absorption is
in general described by a density matrix rather than a wave function; as the degrees of freedom corresponding to escaping particles are
``integrated out'', some coherence effects are lost, see also e.g. \cite{Greenman2010}.

We will assume that the system starts out in a resonance
state and decay by emitting a single particle.
Of course it may rightfully be asked what is meant by ``starting out in a resonance state'' as such states are not physical. We will not enter into this discussion here, however, but rather simply exploit the fact that the resonances may be represented as square integrable eigen states of some (artificial) non-Hermitian Hamiltonian.
We will assume that after emission of a particle, several final states are accessible.
In this context, only Hamiltonians with no
time dependence are considered. As it turns out,
all time dependence is seen to follow the exponential decay law, and the resulting formulae for partial widths are time independent.

The paper is organized as follows: In Sec.~\ref{CAPsection}, the formalism for a CAP is presented. For completeness, the results of \cite{Selsto2010} are briefly outlined. In Sec.~\ref{RateEqForCAP} it is demonstrated how the Lindblad equation reproduces the corresponding classical rate equations, and from this correspondence the partial widths are identified. The loss of coherence is briefly discussed in Sec.~\ref{LossOfCoherence}. In Sec.~\ref{SecECS} the formalism is generalized so that it may also be applied to obtain partial widths using exterior complex scaling (ECS). In Sec.~\ref{Disc} the applicability of the proposed formulae is discussed, and some remaining challenges are pointed out. Conclusions are drawn in Sec.~\ref{Conclusion}.

\section{Particle loss due to a complex absorbing potential}
\label{CAPsection}

Any non-Hermitian Hamiltonian may be written as the sum of a Hermitian and an anti-Hermitian part;
\begin{equation}
\label{Hamiltonian}
H= H^\mathrm{h} - i H^\mathrm{ah} \quad \text{ with } \quad  \quad H^\mathrm{h} = (H^\mathrm{h})^\dagger \quad \text{ and }
\quad H^\mathrm{ah} = (H^\mathrm{ah})^\dagger.
\end{equation}
We will assume that all eigenvalues of $H$ have non-positive imaginary
parts. A sufficient condition for this is that
$H^\mathrm{ah}$ be positive semi-definite \cite{Bendixson1902,Wielandt1955}.
In \cite{Selsto2010} the anti-Hermitian part consisted in a CAP. It was
demonstrated, expressing all interactions in terms of second quantization, that
the Lindblad equation, contrary to the Schrödinger equation, is able to restore the dynamics of the remaining particles
after absorption of other particles in a consistent manner.

Suppose an $N$-particle system is described on a numerical grid $\{ x_i\}$, where
$x_i$ refers to the position, and also other degrees of freedom such as the spin of
a particle. Here, a countable representation has been chosen for notational simplicity.
With creation and annihilation operators $c_i^\dagger$ and $c_i$,
where $c_i^\dagger$ creates a particle at ``position'' $x_i$ and $c_i$ annihilates a
particle at $x_i$, the Hamiltonian may be expressed in a manner which
does not depend on the number of particles at hand:
\begin{equation}
\label{2ndQuantHam}
H=\sum_{kl} h_{k,l} c_k^\dagger c_l + \frac{1}{2}
\sum_{pqrs}V_{pq,rs} c_p^\dagger c_q^\dagger c_s c_r \quad .
\end{equation}
We have here assumed at most two-particle interactions. The field operators obey
the usual anti-commutation relations
\begin{equation}
\label{Kommutator} \{ c_i, c_j \}=0 , \quad  \{ c_i^\dagger, c_j^\dagger \}=0 , \quad  \{ c_i, c_j^\dagger \} = \delta_{i,j}~.
\end{equation}

The CAP, which is a local imaginary potential $-i \Gamma(x)$ with $\Gamma \geq 0$, may be represented analogously by
\begin{equation}
\label{CAPoperator}
\hat{\Gamma} = \sum_i \Gamma(x_i) \, c_i^\dagger c_i.
\end{equation}

Since the process we wish to describe should preserve total probability (trace), $\mathrm{Tr} \rho = 1 \; \forall t$,
and positivity, $\rho \geq 0 \; \forall t$, and the absorption process should be Markovian, our equation of motion should be on Lindblad form \cite{Lindblad1976,Gorini1976}:
\begin{equation}
\label{Lindblad}
i \hbar \dot{\rho} = [\hat{H},\rho] - i \sum_{kl} \gamma_{k,l}
\left(\{A_k^\dagger A_l, \rho\}
- 2 A_l \rho A_k^\dagger \right) \quad .
\end{equation}
By comparison with the von Neumann equation for a non-Hermitian Hamiltonain,
\begin{equation}
\label{VonNeumann}
i \hbar \dot{\rho} = H \, \rho - \rho \, H^\dagger = [H^\mathrm{h}, \rho ] - i \{H^\mathrm{ah}, \rho \} \quad ,
\end{equation}
it is seen that $\hat{H}$ in Eq.~(\ref{Lindblad}) should be identified with the Hermitian part of the Hamiltonian, $H^\mathrm{h}$, and the Lindblad operators should fulfill
\begin{equation}
\label{LindbladAH}
\sum_{k l} \gamma_{k,l} \, A_k^\dagger A_l = H^\mathrm{ah} = \hat{\Gamma} \quad .
\end{equation}
This leads to the equation
\begin{equation}
\label{ResultJPhysB}
i \hbar \dot{\rho}= \left[H, \rho \right] - i \left\{\hat{\Gamma}, \rho
\right\}
+ 2 i \sum_k \Gamma(x_k) \, c_k \, \rho \, c_k^\dagger \quad
\end{equation}
for the density matrix of the entire system.
In this Fock space description, the total density matrix $\rho$ does not correspond to a fixed number of particles.
For an initial state with a well-defined particle number $N$, the total density matrix remains block diagonal, where each block corresponds to a sub-system consisting of $n \leq N$ particles. The $n$-particle sub-system evolves according to
 \begin{equation}
\label{ResultJPhysBorbit_nPart}
i \hbar \dot{\rho}_n =
\left[H, \rho_n \right] - i \left\{\hat{\Gamma}, \rho_n
\right\}
+ i \hbar  \mathcal{S}[\rho_{n+1}] \quad ,
\end{equation}
where the source term
\begin{equation}
\label{SourceTermCAP1}
\mathcal{S} [\rho_{n+1}] = \frac{2}{\hbar} \sum_k \Gamma(x_k) \, c_k \, \rho_{n+1} \, c_k^\dagger \quad .
\end{equation}
For the initial $N$-particle sub-system, the von Neumann
equation is reproduced with the
``effective Hamiltonian'' $H_\mathrm{CAP} \equiv H-i \hat{\Gamma}$,
\begin{equation}
\label{VonNeumannCAP}
i \hbar \dot{\rho}_N = H_\mathrm{CAP} \, \rho_N - \rho_N \, H_\mathrm{CAP}^\dagger \quad ,
\end{equation}
which is equivalent to the Schrödinger equation if the initial state is a pure state, $\rho(t=0) = |\Psi^{(N)} \rangle \langle \Psi^{(N)} |$.

Of course, the field operators $c_i^{(\dagger)}$ could also refer to single particle orbitals, $\chi_i(x)$, rather than points on a grid. In such a representation, the CAP is in general not diagonal, and
Eq.~(\ref{SourceTermCAP1}) is rewritten as
\begin{equation}
\label{SourceTermCAP2}
\mathcal{S} [\rho_{n+1}] = \frac{2}{\hbar} \sum_{kl} \langle \chi_k| \Gamma | \chi_l \rangle
\, c_l \, \rho_{n+1} \, c_k^\dagger \quad .
\end{equation}

We would like to stress that this formalism is based solely on the Lindblad equation.
The fact that we wish to describe a Markovian process in a manner which preserves trace and positivity, cf. \cite{Lindblad1976,Gorini1976}, is enough to justify the generic form of Eq.~(\ref{Lindblad}). We do not make any explicit reference to any reservoir degrees of freedom to be traced over, nor to any Born-Markov approximation \cite{Breuer2002}.

\subsection{Correspondence with classical rate equations}
\label{RateEqForCAP}

As master equations constitute a way of introducing classical concepts into a quantum mechanical context, is seems justified to hope that the above formalism is able to reproduce the corresponding rate equations for the decay process. Note, however, that there is no {\it a priori} guarantee that rate equations are relevant here; we are dealing with a quantum mechanical process in which not only probabilities, but also coherence effects may play a role.
To the extent that the population of the various states involved do follow rate equations, the populations should
fulfill
\begin{eqnarray}
\label{RatesExampleOne1}
\dot{P}_\mathrm{res} & = & -r P_\mathrm{res}
\\
\label{RatesExampleOne2}
\dot{P}_p & = & + r_p P_\mathrm{res} \quad ,
\end{eqnarray}
which, with initial conditions $P_\mathrm{res}(t=0)=1, \; P_p(t=0)=0 \; \forall \, p$,
have the solutions
\begin{equation}
\label{SolutionPop}
P_\mathrm{res} = e^{-r t}, \quad P_p=\frac{r_p}{r} (1 - e^{-r t}) \quad .
\end{equation}
Here $P_\mathrm{res}(t)$ is the population of the initial meta-stable state, and $P_p(t)$ is the population of the $p$-th decay product, which is assumed to be stable.
In order for total population to be conserved, the partial rates $r_p$
should sum up to the total rate;
\begin{equation}
\label{SumToTotal}
\sum_{p} r_p = r.
\end{equation}
The (partial) rates $r_{(p)}$ are directly related to the (partial) {\it widths} $\Gamma_{(p)}$ as they simply differ by a factor $\hbar$; the terms ``rates'' and ``widths'' are often used interchangeably.

We may use the complex spectrum of $H_\mathrm{CAP}$ to
identify resonances \cite{Riss1993,Santra2006}. We will take an $N$-particle resonance state $\psi_\mathrm{res}^{(N)}$ with
\[
H_ \mathrm{CAP} | \psi_\mathrm{res}^{(N)} \rangle = \varepsilon_\mathrm{res}^{(N)} | \psi_\mathrm{res}^{(N)} \rangle \quad
\]
to be our initial state, $\rho(t=0)= |\psi_\mathrm{res}^{(N)} \rangle \langle \psi_\mathrm{res}^{(N)}|$. We have also assumed that the system is stable after emission of a
single particle, i.e., the state $\psi_\mathrm{res}^{(N)}$ is such that asymptotically only one particle may be found. As the particle belonging to the single particle continuum escapes, it is absorbed, and the
$(N-1)$-particle state $\rho_{N-1}$ is reconstructed via the source term, cf. Eqs.~(\ref{ResultJPhysBorbit_nPart},\ref{SourceTermCAP1}).
We will denote the eigen states of the $(N-1)$-particle system by $\varphi_p^{(N-1)}$. As the only unbound particle is removed, only bound $(N-1)$-particle states may be populated. (This may be seen from Eqs.~(\ref{ResultJPhysBorbit_nPart},\ref{SourceTermCAP1}).)
Since these states have a negligible overlap with the CAP, they all have real eigen energies, and they are all orthogonal to each other;
\begin{eqnarray}
\nonumber
H_\mathrm{CAP} | \varphi_p^{(N-1)} \rangle & = & H | \varphi_p^{(N-1)} \rangle = \varepsilon_p^{(N-1)} | \varphi_p^{(N-1)} \rangle \\
\label{RealEnergy}
\mathrm{Im} \, \varepsilon_p^{(N-1)} & = & 0 \quad \forall \, p \\
\nonumber
\langle \varphi_p^{(N-1)} | \varphi_q^{(N-1)} \rangle & = & \delta_{p,q} \quad .
\end{eqnarray}
The eigen energy of the initial state, however, is complex, and its imaginary part is negative
\begin{equation}
\label{ComplexEnergy}
\mathrm{Im} \, \varepsilon_\mathrm{res}^{(N)} < 0, \quad \text{ defining } \quad \varepsilon_\mathrm{res}^I \equiv - \mathrm{Im} \, \varepsilon_\mathrm{res}^{(N)} \quad .
\end{equation}

The population of the initial state $P_\mathrm{res}$ is found from the density matrix as
$\langle \psi^{(N)}_\mathrm{res}|\rho | \psi_\mathrm{res}^{(N)} \rangle =
\langle \psi^{(N)}_\mathrm{res}|\rho_N | \psi_\mathrm{res}^{(N)} \rangle$. With this, Eq.~(\ref{VonNeumannCAP}) provides
\begin{eqnarray*}
&&
i \hbar \dot{P}_\mathrm{res} =
\varepsilon_\mathrm{res}^{(N)} P_\mathrm{res} - P_\mathrm{res} \left( \varepsilon_\mathrm{res}^{(N)}\right)^* =
-2 i \varepsilon_\mathrm{res}^I P_\mathrm{res} \quad \Leftrightarrow
\\
&&
\dot{P}_\mathrm{res} = -\frac{2}{\hbar} \varepsilon_\mathrm{res}^I P_\mathrm{res}
\quad .
\end{eqnarray*}
I.e., Eq.~(\ref{RatesExampleOne1}) is reproduced with the familiar identification $r=2/\hbar \; \varepsilon_\mathrm{res}^I$, or in terms of a width:
\[\Gamma = 2 \, \varepsilon_\mathrm{res}^I \quad . \]
The explicit solution of Eq.~(\ref{VonNeumannCAP}) with the proper initial condition is
\begin{equation}
\label{SolutionNpart}
\rho_N(t) = e^{-rt} \, |\psi_\mathrm{res}^{(N)} \rangle \langle \psi_\mathrm{res}^{(N)}|  \quad .
\end{equation}

The population of the $p$-th $(N-1)$-particle eigen state $\varphi_p^{(N-1)}$ is
\[
P_p(t) = \mathrm{Tr} \left[ |\varphi_p^{(N-1)} \rangle \langle \varphi_p^{(N-1)} | \rho(t) \right] = \langle \varphi_p^{(N-1)} | \rho_{N-1} | \varphi_k^{(N-1)}  \rangle \quad .
\]
With this, Eqs.~(\ref{ResultJPhysBorbit_nPart},\ref{SourceTermCAP2}) and the fact that all the involved eigen energies are real, Eq.~(\ref{RealEnergy}), we have
\begin{eqnarray*}
i \hbar \dot{P}_p & = & \varepsilon_p^{(N-1)} P_p - P_p \left( \varepsilon_p^{(N-1)} \right)^* +
i \hbar \, \langle \varphi_p^{(N-1)} |\mathcal{S}[\rho_N] | \varphi_p^{(N-1)} \rangle \quad \Leftrightarrow
\\
\dot{P}_p & = & \langle \varphi_p^{(N-1)} |\mathcal{S}[\rho_N] | \varphi_p^{(N-1)} \rangle =
e^{-rt}\langle \varphi_p^{(N-1)} |\mathcal{S}\left[ |\psi_\mathrm{res}^{(N)} \rangle \langle \psi_\mathrm{res}^{(N)} |  \right] | \varphi_p^{(N-1)} \rangle = \\
&&
\frac{2}{\hbar} \sum_{kl} \langle \chi_k|\Gamma|\chi_l \rangle \langle \varphi_p^{(N-1)} | c_l  |\psi_\mathrm{res}^{(N)} \rangle \langle \psi_\mathrm{res}^{(N)} | c_k^\dagger | \varphi_p^{(N-1)} \rangle \; P_\mathrm{res}
\quad .
\end{eqnarray*}
We have here used Eq.~(\ref{SourceTermCAP2}).
Eq.~(\ref{RatesExampleOne2}) is reproduced by identifying $r_p$ with $\langle \varphi_p^{(N-1)} | \mathcal{S}[| \psi_\mathrm{res}^{(N)} \rangle \langle \psi_\mathrm{res}^{(N)}|]|\varphi_p^{(N-1)} \rangle$. Thus, the partial width for the $p$-th decay product is \footnote{Admittedly, the notation is somewhat confusing here; on the left hand side ``$\Gamma$'' refers to the partial width, whereas it is the CAP on the right hand side.}:
\begin{equation}
\label{MainResultCAP}
\Gamma_p =  2\,  \sum_{kl} \langle \chi_k|\Gamma|\chi_l \rangle \, \langle \varphi_p^{(N-1)} | c_l |\psi_\mathrm{res}^{(N)} \rangle \, \langle \psi_\mathrm{res}^{(N)} | c_k^\dagger | \varphi_p^{(N-1)} \rangle \quad .
\end{equation}

Partial widths defined as in Eq.~(\ref{MainResultCAP}) will manifestly sum up to the total width, cf. Eq.~(\ref{SumToTotal}). This is ensured by the fact that the Lindblad equation is trace conserving and that the relevant $(N-1)$-particle eigen states are orthogonal, cf. Eq.~(\ref{RealEnergy}). Also the reality and positivity of the partial widths is ensured by the the fact that they are derived from an equation of Lindblad form. This is also clearly seen from a diagonal representation, i.e. a grid representation, of the CAP, Eq.~(\ref{SourceTermCAP1}):
\begin{equation}
\label{MainResultCAPdiag}
\Gamma_p = 2 \, \sum_{k} \Gamma(x_k) \left| \langle \varphi_p^{(N-1)} | c_k | \psi_\mathrm{res}^{(N)} \rangle \right|^2 \quad .
\end{equation}

Conceptually, Eq.~(\ref{MainResultCAP}) offers a rather appealing way of calculating partial widths as there exists such a clear correspondence between the presented formalism and the classical rate equations for the populations.

\subsection{Loss of coherence}
\label{LossOfCoherence}

As information of the escaping particle is removed by the CAP, also the description of the system in terms of a pure state is lost. This loss of coherence may be quantified by the von Neumann entropy, $S(\rho)=-\mathrm{Tr}[\rho \ln \rho]$, or the purity $\varsigma(\rho)=\mathrm{Tr} \rho^2$. For our analysis we find that the latter quantity, which is unity for a pure state and decreasing with the degree of ``mixedness'', will be the most convenient one. With our density operator expressed as $\rho=\rho_N+\rho_{N-1}$, where $\rho_N$ is provided by Eq.~(\ref{SolutionNpart}) and $\rho_{N-1}$ is expressed in terms of the (stable) eigen states,
\begin{equation}
\label{Expansion}
\rho_{N-1} = \sum_{r,s} p^{(N-1)}_{rs}(t) | \varphi_r^{(N)} \rangle \langle \varphi_s^{(N-1)} | \quad ,
\end{equation}
the purity may be expressed as
\begin{equation}
\label{PurityExpression}
\varsigma(\rho) = e^{-2 \Gamma t/\hbar} + \sum_{rs} |p^{(N-1)}_{rs}|^2 \quad .
\end{equation}
The coefficients $p^{(N-1)}_{rs}$ may be obtained from Eq.~(\ref{ResultJPhysBorbit_nPart}) e.g. by means of the Laplace transform. With the proper initial conditions, the solution is
\begin{eqnarray}
\label{CoeffSolution}
p^{(N-1)}_{rs} & = & \frac{\kappa_{rs}}{\Gamma - i \Delta \varepsilon_{rs}} (e^{-i \Delta \varepsilon_{rs} \, t/\hbar} - e^{-\Gamma t/\hbar}), \quad \text{ with} \\
\nonumber
\Delta \varepsilon_{rs} & \equiv & \varepsilon^{(N-1)}_r - \varepsilon^{(N-1)}_s \in \mathbb{R} \quad \text{ and} \\
\nonumber
\kappa_{rs} & \equiv & 2\sum_{kl} \langle \chi_k | \Gamma | \chi_l \rangle \langle \varphi_r^{(N-1)} | c_l | \psi_\mathrm{res}^{(N)} \rangle
\langle \psi_\mathrm{res}^{(N)}| c_k^\dagger | \varphi_s^{(N-1)} \rangle \quad .
\end{eqnarray}
The diagonal elements, or populations, $p_{rr}$ assumes the simpler form provided in Eq.~(\ref{SolutionPop}) ($\kappa_{rr} = \Gamma_r$).
Thus, the purity, as a function of time, is provided by
\begin{equation}
\label{PurityAvT}
\varsigma(\rho) =  e^{-\Gamma t/\hbar} +  \sum_{rs} \frac{\left| \kappa_{rs} \right|^2}{\Gamma^2+\Delta \varepsilon_{rs}^2}
\left( 1 + e^{-2 \Gamma t/\hbar} - 2 \cos(\Delta \varepsilon_{rs} t/\hbar)e^{-\Gamma t/\hbar} \right) \xrightarrow[t \rightarrow \infty]{}  \sum_{rs} \frac{\left| \kappa_{rs} \right|^2}{\Gamma^2+\Delta \varepsilon_{rs}^2}  \leq 1\quad ,
\end{equation}
where equality applies when there is a single final state accessible only.

\section{Partial widths in the context of exterior complex scaling}
\label{SecECS}

As mentioned, there exist several other techniques for obtaining normalizable representations of resonance wave functions.
A rather straight
forward generalization of the above formalism can accommodate for exterior complex scaling. Exterior
complex scaling is introduced by modifying the spatial variable ${\bf r}$ such that the radial distance from the origin, $r$, is rotated by an angle $\theta$ into the first
quadrant of the complex plane for positions beyond a certain radial distance $R_0$, i.e. ${\bf r} \rightarrow R({\bf r})$ such that
\begin{equation}
\label{ECSdef}
|R({\bf r})| = \left\{ \begin{array}{ll} r, & r \leq R_0 \\  R_0 + e^{i \theta}(r-R_0), & r > R_0 \end{array} \right. ~.
\end{equation}
Just as in the case of a CAP, stable states, i.e. localized states, are virtually unaffected by the scaling as long as the ``unscaled region'' is large enough, i.e. as long as $R_0$ is larger than the extension of all populated bound states. With an adequate implementation of ECS, distinguishing resonance states from other continuum states is rather straight forward.
``Ordinary'' continuum states are typically rotated by the angle $2\theta$ into the fourth quadrant of the complex plane from their respective thresholds, whereas resonance states are virtually $\theta$-independent and hence easily separated from the other continuum states given large enough $\theta$.

By extending the formalism for CAPs more or less directly to Hamiltonians modified by ECS, we may arrive at a formula for partial widths valid for this case. If the unscaled Hamiltonian is represented in terms of single particle orbitals, $\chi_i$,
ECS will introduce an anti-Hermitian term in the Hamiltonian which reads
\begin{equation}
\label{ECS2ndQuandAH}
H^\mathrm{ah} = \sum_{kl} h^\mathrm{I}_{k,l} c_k^\dagger c_l + \frac{1}{2} \sum_{pqrs} V^\mathrm{I}_{pq, rs} c_p^\dagger c_q^\dagger c_s c_r
\end{equation}
where
\begin{eqnarray}
\label{OnePartAHcoeff}
h^\mathrm{I}_{k,l} & = & \int_{r>R_0} d^3{\bf r} \,
\left(\chi_k({\bf r}) \right)^* \left[-\sin(2 \theta)\frac{\hbar^2}{2 m} \nabla^2 +\mathrm{Im} \, V_1(R({\bf r})) \right] \chi_l({\bf r}) \quad \text{and}\\
\label{TwoPartAHcoeff}
V^\mathrm{I}_{pq, rs} & = & \int_{r>R_0} \int_{r'>R_0} d^3{\bf r} \, d^3{\bf r}' \,
\left( \chi_p({\bf r}) \right)^* \left( \chi_q({\bf r}') \right)^* \, \left[ \mathrm{Im} \,V_2(R({\bf r}), R({\bf r}')) \right ] \,
\chi_r({\bf r}) \chi_s({\bf r}') \quad .
\end{eqnarray}
Here $V_1({\bf r})$ is the local one-particle potential and $V_2({\bf r},{\bf r}')$ is the interaction.
Of course, the field operators may still also refer to grid-points. In such a context, care must be taken when constructing the coefficients corresponding to the kinetic energy operator from some (high-order) finite difference scheme such that the cusp condition at $|{\bf r}|=R_0$ is fulfilled.

Reasoning completely analogous to what was performed in the Sec.~\ref{CAPsection} leads to the following equation for the evolution of the $n$-particle system:
\begin{eqnarray}
\label{ResultJPhysBnPartECS}
i \hbar \dot{\rho}_n & = &
\left[ H^\mathrm{h}, \rho_n \right] - i\left\{ H^\mathrm{ah}, \rho_n \right\}
+ i \hbar \mathcal{S}_1[\rho_{n+1}] + i \hbar \mathcal{S}_2[\rho_{n+2}] \quad \text{ with }\\
\label{SourceTermECS1}
\mathcal{S}_1[\rho_{n+1}] & = & \frac{2}{\hbar} \sum_{kl} h^\mathrm{I}_{k.l} c_l \, \rho_{n+1} \, c_k^\dagger
\quad \text{ and } \\
\label{SourceTermECS2}
\mathcal{S}_2[\rho_{n+2}] & = & \frac{1}{\hbar} \sum_{pqrs} V^\mathrm{I}_{pq,rs} c_s c_r \, \rho_{n+2} \, c_p^\dagger c_q^\dagger \quad .
\end{eqnarray}
If the initial system is represented by an $N$-particle pure state $\Psi^{(N)}$, we have the following special cases:
\begin{eqnarray*}
\rho_N &  = &  | \Psi^{(N)} \rangle \langle \Psi^{(N)} | \quad \text{ with } \quad
i \hbar \dot{\Psi}^{(N)} = H \Psi^{(N)} \quad \text{ and } \\
i \hbar \rho_{N-1} & = & H \rho_{N-1} - \rho_{N-1} H^\dagger + i \hbar \mathcal{S}_1\left[ |\Psi_N \rangle \langle \Psi_N| \right] \quad .
\end{eqnarray*}

Arguably, the biggest difference between this approach and the CAP-approach is the fact that Eq.~(\ref{ResultJPhysBnPartECS}) features a two-particle source term, as opposed to Eq.~(\ref{ResultJPhysBorbit_nPart}) which only includes a one-particle source term. The ``flow'' is still diagonal, however; there are no elements in the total density matrix which correspond to a particle number which is not well defined (unless such states were populated initially).
If we again are to consider a resonance state which is stable after emission of a particle, the two-particle source term is not of crucial importance, however. The initial ``stationary'' state $\psi_\mathrm{res}^{(N)}$ will have at most one particle in the region beyond $R_0$ if $R_0$ is chosen large enough. Since $\psi_\mathrm{res}^{(N)}$ is an eigen state of the non-Hermitian Hamiltonian, this situation will remain. Thus, as $\mathcal{S}_2[\rho_N]$ removes two particles in the region beyond $R_0$, its contribution vanishes in this situation.

In the same manner as above, the classical rate equations, Eqs.~(\ref{RatesExampleOne1},\ref{RatesExampleOne2}), are reproduced by Eq.~(\ref{ResultJPhysBnPartECS}). The partial rate $r_p=\Gamma_p/ \hbar$, i.e. the rate at which the $p$-th eigen state of the $(N-1)$-particle system is populated, is now calculated as
\begin{equation}
\label{MainResultECS}
\Gamma_p = 2 \, \sum_{k,l} h^\mathrm{I}_{k,l} \, \langle \varphi_p^{(N-1)}| c_l | \psi_\mathrm{res}^{(N)} \rangle \, \langle \psi_\mathrm{res}^{(N)}| c_k^\dagger | \varphi_p^{(N)} \rangle \quad ,
\end{equation}
where $h^\mathrm{I}_{k,l}$ is provided by Eq.~(\ref{OnePartAHcoeff}). Technically, the only difference between the above formula and Eq.~(\ref{MainResultCAP}) lies in how the coefficients $\langle \chi_k|\Gamma| \chi_l \rangle$, resp. $h^\mathrm{I}_{k,l}$, are obtained.

\section{Discussion}
\label{Disc}

Let us first address the advantages with the formalism presented here. The greatest numerical advantage lies, as in \cite{Selsto2010}, in the fact that rather than analyzing an unbound $N$-particle wave function, we may now obtain the relevant information by studying a localized $(N-1)$-particle system -- once the resonance state is obtained. This is a considerable simplification for two reasons. First, the ``curse of dimensionality'' is reduced. Second, partial widths are obtained without explicit reference to scattering states; the finite set of {\it bound states} suffices once the initial resonance wave function has been determined. The fact that the population dynamics is dictated by an equation of Lindblad form ensures that all partial widths sum up to the total width.

For the predictions of Eq.~(\ref{MainResultCAP}) or Eq.~(\ref{MainResultECS}) to be reliable, the initial resonance state must be well represented. The complex resonance energy should remain constant as the absorber region is moved outwards. Moreover, the change induced by a reduction in the strength of the CAP or, for ECS, an increase in the scaling angle $\theta$ should be minimal.
It must be checked in any implementation to what extent these criteria are met.
Clearly, the partial widths $\Gamma_p$ are subject to the same criteria for invariance as the total width $\Gamma$.

Some issues remain unresolved. Possibly the most obvious one is that no formalism has been presented for (uniform) complex scaling, which is imposed simply by multiplying the position vector by a complex phase,
${\bf r} \rightarrow {\bf r} e^{i \theta}$.
A generalization of the present formulae which also accommodates for this form of non-Hermicity is desirable. This is not trivially obtained, however.
Another rather unsatisfactory feature is revealed if we consider an initial state which is not a single resonant state. Supposing our initial state may be written as the super position $\Psi^{(N)} = c_1 \varphi_1^{(N)} + c_2 \varphi_2^{(N)}$. If $\varphi^{(N)}_{1,2}$ are eigen states, it would seem natural to interpret the incoherent sum $|c_1|^2+|c_2|^2$ as the total population of the initial $N$-particle system. However, as the states are eigen states of a non-Hermitian Hamiltonian, they will in general not be orthogonal, and the sum $|c_1|^2+|c_2|^2$ will not provide the trace of $\rho_N$. Thus, it is not straight forward to interpret such a state in terms of populations.

Another interesting question arises if we omit the assumption that the system is stable after emission of a single particle, i.e. the initial resonance state does not exclusively belong to the single-particle continuum. Can we still use these concepts to describe unstable systems which decay into sub-systems which themselves may be unstable? A straight forward application would not provide reasonable results for the same reason as above; the wave functions of the intermediate decay products, $\varphi_p^{(N-1)}$, would in general not be orthogonal ($\mathrm{Im} \, \varepsilon_p^{(N-1)}<0$).

All these issues call for a generalization of the formalism presented here. It seems that such a generalization should be based on a bi-orthogonal basis representation rather than an orthogonal one \cite{Elander1989}.

\section{Conclusion}
\label{Conclusion}

Formulae for calculating partial widths for decay-processes in which the remaining system is stable after emission of one particle was derived. The formalism combined complex absorbing potentials and exterior complex scaling, respectively, with second quantization and the Lindblad equation. The partial widths was derived in a consistent manner which ensured that they sum up to the total width of the original meta-stable state. The proposed way of calculating partial widths should be rather easily implemented once a good representation of the initial resonance state is obtained, and it is also believed to be numerically favorable.

\section{Acknowledgements}

Discussions with dr. Simen Kvaal and dr. Michael Genkin have been quite beneficial to this work. It has also benefited from inspiring interactions with prof. Eva Lindroth and with prof. Nimrod Moiseyev.

\end{document}